\documentclass[12pt,a4paper]{article}
\usepackage{amssymb,amsfonts,amsmath}
\usepackage{color}
\usepackage{epsfig, euscript, graphics}

\title{Quantum postulate vs. quantum nonlocality: Is Devil  in $h?$}

\author{Andrei Khrennikov\\ 
Linnaeus University, International Center for Mathematical Modeling\\  in Physics and Cognitive Sciences
 V\"axj\"o, SE-351 95, Sweden}

\begin{document}
\maketitle
 
\abstract{This note is a part of my efforts for getting rid of nonlocality from quantum mechanics (QM).  Quantum nonlocality is two faced Janus, one face is L\"uders projection nonlocality, another face is Bell nonlocality. This paper  is devoted to disillusion of the latter. The main  casualty of Bell's  model with hidden variables is that it straightforwardly contradicts to the Heinsenberg's uncertainty and generally Bohr's complementarity principles. Thus, we do not criticize the derivation or interpretation of the Bell inequality (as was done by numerous authors). Our critique is directed against the model as it is.  
The original Einstein-Podolsky-Rosen (EPR) argument was based on the Heinseberg's principle, but EPR did not question it. Hence, the arguments of EPR and Bell differ crucially.    It is worth to find the physical seed of the aforementioned principles. This is the  {\it quantum postulate}: the existence of indivisible 
quantum of action.  Bell's approach with hidden variable straightforwardly implies rejection of the quantum postulate. 
Heisenberg compared the quantum  postulate  with constancy of light's velocity in special relativity. Thus attempts to explain long distance correlations within  the Bell model can be compared with attempts to construct models violating the laws 
of relativity theory. Following Zeilinger, I search for the fundamental principles of quantum mechanics (QM) similar to the principles of relativity and consider the quantum action and complementarity principles as such principles.}

\section{Introduction}

Recently I published a series of papers which can be unified by the slogan ``getting rid of nonlocality from quantum physics'' 
\cite{NL1}-\cite{NL2}. The wide use of the notion of quantum nonlocality overshadows the real output of quantum theory, mystifies it,  generates unjustified expectations and speculative interviews for mass-media of otherwise very respectable scientists.   The aim is to decouple nonlocality from quantum theory. The main message of aforementioned papers is that quantum theory is local, that 
``spooky action at a distance'' was just shicky Einstein's slogan \cite{Boughn1} from a letter to Born in 1947   \cite{EB}. Einstein directed it against {\it the individual interpretation} of a quantum state. This interpretation is often referred as {\it the Copenhagen interpretation} of QM. From his viewpoint, one should either reject  this interpretation or confront with spooky action at a distance 
(see paper \cite{NL3} for details and the probabilistic analysis).  
This viewpoint was especially clearly presented in letters' exchange between Einstein and Schr\"odinger \cite{ESCH}. 

One of complications in getting rid of nonlocality from QM is that so-called ``quantum nonlocality'' is two faced Janus
\cite{NL3} (see also Appendix 1). People freely refer to his different faces, mix them, and often cannot distinguish them.  Two faces of nonlocality Janus are 
\begin{itemize}
\item  {\it L\"uders nonlocalty:} apparent  nonlocality of QM based on the projection postulate and discussed in the EPR-paper \cite{EPR}\footnote{In fact, the most consistent representation of this nonlocality can be found in Aspect's papers \cite{AA0,AA1}. Aspect did not refer to EPR ``elements of reality'' and he proceeded straightforwardly with the L\"uders projection postulate \cite{Luders}. We remark that the projection postulate is often referred as the von Neumann- L\"uders postulate or even simply the von Neumann postulate. However, von Neumann  \cite{VN} sharply distinguished the cases of observables with non-degenerate and  degenerate spectra. For the later case, he considered a more general form of the state transformation generated by observation back-action; in particular, by von Neumann a pure initial state can be transferred into a mixed state, as in the modern theory of quantum instruments \cite{DV,Oz1}. EPR used the projection postulate for arbitrary observables, i.e., as was later formalized by L\"uders \cite{Luders}.};
\item {\it Bell nonlocality:} subquantum nonlocality based on misleading interpretation of violation of the Bell inequalities
 \cite{Bell0}-\cite{CHSH}.
\end{itemize}

Typically by saying ``quantum nonlocality'' one does not specify whether this is L\"uders or Bell nonlocalty (often a debater even does not understand the difference between these nonlocalities). So, the first step to elimination of nonlocality from quantum theory is learning its Janus-like structure  (Appendix 1); see paper \cite{NL3}
 devoted to illuminating  this structure and disillusion of  L\"uders nonlocalty. 

In paper \cite{NL1}, violation of the Bell inequalities was treated in the purely quantum framework, i.e., without coupling to hidden variables (cf. \cite{FS1}-\cite{PGF}, \cite{Boughn1}). What does quantum theory say about (non)violation of the Bell inequalities? 
In this framework, violation vs. satisfaction of such inequalities  is equivalent to 
{\it local incompatibility vs. compatibility of observables.} These inequalities shouls be treated as statistical tests 
for the complementarity principle (see section \ref{PC}).

However, one can say that the genuine quantum viewpoint on the Bell inequalities is not interesting. The essence of these inequalities 
is in their derivation on the basis of the Bell model with hidden variables  \cite{Bell0}-\cite{CHSH}. In this paper, we want to terminate this line of thinking by showing that 
 the Bell hidden variables project  is in striking contradiction with  quantum foundations. To see this, one need not to derive any inequality. From the very beginning (already by setting the hidden variable model), it is clear that {\it Bell's model confronts with  the fundamental principle of QM, the Bohr complementarity principle} (see Bohr \cite{BR} and also \cite{PL1,PL2,F}) and, in particular, the Heisenberg uncertainty principle. Thus, now we do not criticize the derivations and interpretations of the Bell type inequalities (cf. \cite{Cetto0}-\cite{KHRB3}). We stress that by starting with the Bell's hidden variables one goes to a Crusade against complementarity.

If one accepts this viewpoint, then the following natural question immediately arises: Why should the complementarity principle be violated only for compound systems?
(If it were violated.)  It seems that there is no reason for this. And studies on intrasystem entanglement (of the degrees of 
freedom of say a single atom)  and, in particular, classical entanglement (of the degrees of 
freedom of say classical electromagnetic field) confirm  this viewpoint (see \cite{CLE} for a review and \cite{NL2} for coupling of 
classical entanglement and complementarity).   

It is often said that the aim of the Bell hidden variables project was explanation of the long distance correlations (see Appendix 2); we discuss this question in section \ref{VSD}. We stress that Bell's  attempt of explanation is based on the rejection of the complementarity principle. It seems that the price of such an attempt  is too high.

By struggling with Bell nonlocality and opposing it to the complementarity principle, 
it is worth to find the physical  seed of this principle.
This seed is the Bohr  {\it quantum postulate} \cite{BR1,BR2,BR2a,BR} declaring the existence of indivisible quantum of action (Planck quantum). Thus, resolution of long debates on quantum nonlocality (Appendix 1), action at a distance,  Bell inequalities is possible only on the basis of this postulate. This crucial issue is totally missed in these debates. 

We recall (in section \ref{ph1})  Bohr's and Heinsenberg's viewpoints on the quantum postulate, especially Heinsenberg's comparison of the existence of quantum of action ($h\not=0)$  with finiteness of light's velocity ($c<\infty)$ and its independence of the inertial frame  \cite{BR1,BR2,BR}.

In section \ref{ph1}, we recall the practically forgotten paper of Zeilinger \cite{Z} in that he looked for the fundamental principles of QM similar to the principles of special relativity. The quantum postulate  definitely plays the crucial role in formulation of these  fundamental principles (section \ref{ph1}). This postulate can be considered as the physical seed of the Bohr principle of complementarity. The latter is one of the fundamental principle of QM. But, this principle is about observations, the way of extracting of information about quantum systems. This is an epistemological principle \cite{Harald1,Harald2,PL1,PL2}.\footnote{During his life Bohr presented a variety of versions of the complementarity principle. In papers \cite{NL0B,NL3,GG}, I expressed my vision on Bohr's ideas as the block of  sub-principles (see section \ref{PC}). I think that such a compact formulation of Bohr's principles is important for further discussions of the type ``Bohr vs. Bell'' \cite{NL0B}. Nowadays, the Bohr complementarity principle is discussed only by philosophers, e.g., in \cite{PL1,PL2,F}, in the form of citations followed by long discussions.}  The Bohr's quantum postulate is ontic, it is about physical reality as it is. 

Similarly to Einstein's formulation of the relativity principle on light's velocity \cite{Ref}, we formulate the quantum action principle based of the Bohr quantum postulate (section \ref{ph1}).  Following  von Weizs\"acker \cite{W} and Atmanspacher and Primas \cite{Harald1,Harald2}, we consider QM as an epistemic theory, a theory about knowledge (see also \cite{OE1,OE2}). So, the quantum action principle is the epistemic counterpart  of Bohr's quantum postulate on the existence of indivisible quantum of action, this postulate is of the ontic nature.      

This is a good place to recall  the recent attempts to derive the quantum formalism from ``natural probabilistic and information
principles'',  e.g., in  \cite{DA1}-\cite{Ch2}. This activity differs from our attempt (following Zeilinger \cite{Z}) to formulate 
the fundamental principles of QM. We do not try to derive its mathematical formalism. The latter can be compared with the dressing for 
the main dish, we want to test the dish without dressing.     

We start with the comparative analysis of the views of Einstein, Podolsky, and Rosen  and Bell (sections \ref{EPRV}, \ref{BLV}).
Typically one considers Bell as the follower of EPR and claims that the Bell inequality is straightforwardly  related to the EPR-pradox. It seems that this viewpoint is misleading. Then we point (section \ref{SS}) that by considering Bell's hidden variables model, one 
struggles against the complementarity principle and, in fact, against the existence of the Planck quantum of action. 

Would one like to ``explain'' the long distance correlations by the cost of confronting again with the ultraviolet (Rayleigh–Jeans) catastrophe and rejecting the original Planck work on the black body radiation? (see section \ref{VSD}).

\section{EPR}
\label{EPRV}

The EPR paper \cite{EPR} was directed against the Copenhagen interpretation of the wave function. 
Since this interpretation has many versions (Plotnitsky even proposed to speak about interpretations in the 
spirit of Copenhagen \cite{PL1,PL2}), it is important to specify the EPR treatment of this interpretation. 

{\bf Copenhagen interpretation} (EPR) {\it Wave function (quantum state) $\psi$ represents the state of 
an individual quantum system.}

It is important to stress that ``state'' is interpreted epistemically as representing knowledge about possible outcomes of measurements
on the system in the state $\psi.$ So, $\psi$ is not the ontic state - not the state of the system   as it is, i.e., without relation
to external observations. State's interpretation in the EPR-paper is very close the modern information interpretations used in quantum 
information theory. This point has not been so much highlighted (see, however, \cite{W,Harald1,Harald2,PARX}).    

By this interpretation the quantum mechanical description based on the wave function representation of the state of a quantum system is complete. The complete physical theory  is defined as follows \cite{EPR}: \footnotesize{{\it any element of physical reality have a counterpart in the physical theory.}} 

The EPR-reasoning was based on two basic quantum mechanical principles:
\begin{itemize}
\item reduction of the wave function (the projection postulate) resulting from measurement's back-action; 
\item the Heisenberg uncertainty principle.
\end{itemize}
The latter was formulated as follows: \footnotesize{{\it``It is shown in quantum mechanics that, if the operators corresponding to two physical quantities, say $A$ and $B,$ do not commute, $AB\not=BA,$ then the precise knowledge of one of them precludes such a knowledge of the other. Furthermore, any attempt to determine the latter experimentally will alter the state of the system in such a way as to destroy the knowledge of the first.''}}

EPR showed that  the assumption that QM (endowed with the Copenhagen interpretation) is a complete theory implies violation of the Heisenberg uncertainty principle. Since they were sure in validity of this principle, EPR concluded that the quantum mechanical description of nature is incomplete.

We emphasize that EPR did not question the validity of the Heisenberg principle (see Appendix 3). If it would be possible to violate 
this principle, then assigning  to the same system two wave functions which are eigenfunctions of observables represented by non-commutative operators would not lead  to any problem (Appendix 3).

Thus, by concluding that \footnotesize{{\it``... the wave function does not provide a complete description of the physical reality, we left open the question of whether or not such a description exists'' and believing ``...that such a theory is possible'',}} they do not dream for 
a theory violating the Heisenberg's uncertainty principle.  By reading later works of Einstein we can guess that he wanted to construct
a classical field model underlying QM \cite{EI} (see Appendix 4 for such an attempt). 

We remark that EPR did not question validity of quantum mechanical description, they were just looking for a more detailed description.
But, this deeper description should respect the basic principles of QM, including the uncertainty and complementarity principles.

\section{Bell}
\label{BLV}

Although Bell started his paper \cite{Bell0} with referring to the EPR paper as proving incompleteness of QM, his model with hidden variables has not so much to do with the EPR-dream for a complete physical theory generalizing QM. It is surprising that this inconsistency has never been emphasized in numerous papers on Bell's inequality (see, e.g., Aspect \cite{AS3}). The main difference of 
Bell's model from the EPR-dream is that his model is in the striking contradiction with the quantum mechanical 
description, especially with  the Heisenberg uncertainty principle (see Appendix 3). 

 Consider Bell's random variables
$A(a, \lambda), B(b, \lambda)$  representing observables of Alice and Bob, respectively. Surprisingly,  Bell did not highlighted that, besides probabilities
$p_{a,b}(x,y)= p(A(a, \lambda)=x, B(b, \lambda)=y)$ for compatible observables,  Bell's model describes probabilities
$p_{a,a^\prime}(x_1,x_2)= p(A(a, \lambda)=x_1, A(a, \lambda)=x_2)$ for generally incompatible observables (represented by noncommuting 
operators.\footnote{This problem is especially clear in consideration of CHSH-inequality \cite{CHSH}.} From the very beginning, i.e., without any Bell's type inequality, this assumption contradicts to QM-representation of observables and, hence,  to the Heisenberg uncertainty principle (or more generally  to the Bohr complementarity principle).

Of course, Bell may proceed with his special class of subquantum models,  but {\it  without identification of the values of his random variables with values of quantum observables and without identification of ``hidden correlations'' with the experimental correlations.} (De Broglie emphasized   \cite{DB} this viewpoint.) But, Bell wanted experimental verification...

Thus, from the very beginning Bell's model of hidden variables was designed as contradicting the uncertainty principle.
Therefore, it is not surprising that, as was shown in my recent paper \cite{NL1}, violation-satisfaction of the CHSH-inequality 
can be formulated in terms of noncommutativity-commutativity of operators representing local observables of Alice and Bob, respectively. 

\section{Crusade against complementarity}
\label{SS}

The Heisenberg uncertainty principle was the starting point for Bohr's formulation of the complementarity principle 
\cite{BR1}-\cite{BR2a} (see my recent papers \cite{NL0B,NL1,GG} for non-philosophers gently presentation of this principle, see also  section \ref{PC}). Thus, in the light of above consideration,
we can say that in fact Bell's argument was directed against the Bohr complementarity principle. 
This Crusade against complementarity was overshadowed by nonlocality issue (Appendix 1). 
Of course, it is clear that rejection of complementarity principle (or Heisenberg's uncertainty principle) would have similar 
catastrophic consequences even for non-compound systems, say a single atom or neutron, as we can see from so-called contextuality 
tests (see, e.g., \cite{C1}). 

In short, we can say that to discard the Bell model with hidden variables, there is no need to derive inequalities and test them experimentally (of course, if one believes in the basic principles  of QM.)\footnote{If one does not, she should say explicit about this, about her battle with the quantum postulate.} The main impact of experimental tests \cite{Aspect}-\cite{Shalm} is demonstration 
that quantum correlations (predicted by QM) are preserved for long distances. The latter plays the crucial role in quantum engineering.
However, correlations preservation can be checked directly without inequalities. Moreover, by operating with say CHSH-combination of correlations
experimenter can miss mutual compensation of deviations from QM. In Aspect's pioneer experiment \cite{AAA}, 
correlations did not match the quantum prediction, but they mystically compensated each other to violate the Bell inequality
(see \cite{AD} for discussion). (In spite of numerous discussions with experimenters, I am still not sure that data from the basic 
experiments on say CHSH-inequality is clean from the mentioned Aspect-type anomaly. Papers typically present only the CHSH-correlation
combination, but not separate correlations for pairs of experimental settings.)

\section{Explaining:  long distance correlations vs. violation of complementarity principle}
\label{VSD}

Typically, followers of the Bell argument (that has not so much to do with the original EPR-argument) say they want to explain 
the long distance correlations (see Appendix 2). I think that the essence of the problem is in the word ``explain''. 

In science, we operate with mathematical models of physical processes. So, ``explain''  means ``to describe by some mathematical model''. And  quantum mathematics, as a mathematical model, describes perfectly the long distance correlations: entangled states and projection 
type measurements. So, it seems that Bell  and his followers have something different in mind.

Why Bell was not satisfied with the quantum mechanical description? 

From reading Bell, I have the impression that he  ``simply'' wanted to re-establish realism of classical physics. But, what is the main quantum barrier for  such realism? Everybody knows this very well, this is the Bohr complementarity principle with starting point at the Heisenberg uncertainty relation.\footnote{See Bohr \cite{BR1}, {\it``... an independent reality in the ordinary physical sense can neither be ascribed to the phenomena nor to the agencies of observation.''}} And this is clearly stated in the EPR-paper. So, Bell and his followers have to say: we want to break the Heisenberg uncertainty relations. Unfortunately, it was never stated explicitly. Instead, people operate with such an ambiguous notion as ``local realism''. 

Suppose somebody, say Alice, questions the Heisenberg uncertainty principle. Then, why should she consider compound systems? Does she think that these principle is violated only for compound systems?
It would be really strange. Thus, before trying to explain the long distance correlations with the Bell-type hidden variables model, it would be reasonable to try explain incompatibility of observables corresponding spin projections to different axes or incompatibility of position and momentum observables.

The main feature of the Bell model with hidden variables, the feature crying for justification, is violation of the 
complementarity principle. It is not so natural to try to``explain'' long distance correlations without any attempt 
to explain violation of this principle.  

\section{The root of complementarity: Devil is in the Planck constant}
\label{ph}

Thus, by starting the anti-complementarity battle it is useful to remind the foundational roots of complementarity. The Bohr's complementarity principle will be discussed in detail in section \ref{PC}. 

For Bohr, the root  of the complementarity is the existence of {\it indivisible quantum of action} given by the Planck constant $h.$  The existence of this quantum prevents separation of the genuine physical features of a system from the features of interaction with a measurement apparatus. So, the seed of the Bohr complementarity principle is the Planck constant $h.$

It is meaningless to start a Crusade against complementarity without trying to understand the origin of this fundamental quantum of action in nature. Neither Einstein nor Bell tried to perform such investigation;  in fact, neither Bohr nor Heisenberg, for them this is just the feature of nature such as, e.g.,  the constancy of  light's velocity $c.$ And, for the moment, this position can be considered as the only possible.

\section{Quantum action principle}
\label{ph1}

We recall that Zeilinger was looking for the fundamental principle of QM \cite{Z}, similar to Einstein's principle
of relativity: 

{\it The laws of physics are invariant (i.e. identical) in all inertial frames of reference.}

And he formulated the following principle of  quantization of  information: 

{\it An  elementary system represents the   truth value of one   proposition.}

Surprisingly, in his paper \cite{Z} Zeilinger did not mention the Bohr complementarity principle. In fact, Zeilinger's postulate 
is nothing else than Bohr's statement on quantum phenomenon. The latter can be considered as a part of the complementarity principle
(see, especially, my recent papers \cite{NL2}). We shall be back to this issue in section \ref{PC}. 

Now we recall that theory of special relativity is based on two Einstein's principles, and the second one is about light's velocity:  

{\it The speed of light in a vacuum is the same for all observers, regardless of the motion of the light source or observer.}

(In particular, this principle presumes finiteness of light's velocity.) We now point to the close quantum analog of this principle.

Bohr stressed \cite{BR1} that the essence of quantum theory  \footnotesize{{\it ``may be expressed in the so-called quantum postulate, which attributes to any atomic process an essential discontinuity, or rather individuality, completely foreign to the classical theories and symbolised by Planck's quantum of action.''}} On the basis of  Bohr's quantum postulate, we formulate the following principle of QM that can be considered as the analog of   Einstein's second principle: 

\medskip

{\bf Quantum action principle:} {\it Quantum of action is  the same for all observers, regardless experimental contexts.}

\medskip

We can say that this principle is the epistemic counterpart of the Bohr's quantum postulate. The formulation of the quantum action principle involves observables, but the quantum postulate, the existence in nature of indivisible quantum of action, is about nature as it is, i.e., this is the ontic postulate.  

Nowadays, it is practically forgotten that by formulating the uncertainty principle Heisenberg pointed to   the  analogy 
with the light  velocity constraint in special relativity. This analogy was then emphasized by Bohr \cite{BR1,BR2}. In this paper, Bohr
used the term ``reciprocal uncertainty'' for the Heisenberg uncertainty relation. (This reciprocity is related 
to position and momentum.) 
 
\footnotesize{{\it ``Heisenberg has rightly compared the significance of this law of reciprocal uncertainty for estimating the self-consistency of quantum mechanics with the significance of the impossibility of transmitting signals with a velocity greater than that of light for testing the self-consistency of the theory of relativity. ...  Planck's discovery has brought before us a situation similar to that brought about by the discovery of the finite velocity of light.''}} 

For the formulation of the complementarity principle, the concrete value of the Planck constant is not important. It is important only that this quantum of action exists,  $h\not=0.$  In the same  way, the concrete value of light's velocity is not important 
for formulation of special relativity, i.e., it is only important that it is finite, $c< \infty.$ We also stress that constancy of 
action quantum, its independence of observable (measurement procedure), plays the crucial role  in QM, as well as constancy of light’s velocity in special relativity.

\medskip

What are other principles of quantum theory? We shall discuss this problem in section \ref{PC}.

\section{Bohr's complementarity principle}
\label{PC}

In 1949,  Bohr \cite{BR} presented the essence of complementarity in the following widely citing statement: 

\footnotesize{{\it “This crucial point ...  implies the impossibility of any sharp separation between the behaviour of atomic objects and the interaction with the measuring instruments which serve to define the conditions under which the phenomena appear. In fact, the individuality of the typical quantum effects finds its proper expression in the circumstance that any attempt of subdividing the phenomena will demand a change in the experimental arrangement introducing new possibilities of interaction between objects and measuring instruments which in principle cannot be controlled. Consequently, evidence obtained under different experimental conditions cannot be comprehended within a single picture, but must be regarded as complementary in the sense that only the totality of the phenomena exhausts the possible information about the objects.”}}  

By analyzing this Bohr's statement, I propose \cite{NL0B,NL1,GG}  to present the Bohr complementarity principle as the following five interconnected principles:
\begin{itemize}
\item  {\bf Contextuality:} Irreducible dependence of measurement's output on the experimental context.
\item  {\bf Context complementarity:} Existence of complementary experimental contexts. 
\item  {\bf Individuality:} Discreteness of quantum measurements -generation of physical phenomena.
\item {\bf Completeness:}  Complementary observations  provide complete information 
about system's state. 
\end{itemize}
In this formulation, the complementarity principle can be treated as an epistemological principle (see, especially, paper \cite{GG} on
coupling to quantum  information theory).  

Typically, one identifies the Bohr complementarity principle with 
 {\bf Context complementarity}. However, the above citation implies combination of all four ``sub-principles.'' Besides {\bf Context complementarity},  the principles {\bf Contextuality}  and {\bf Completeness}  also attract some attention, but {\bf Individuality} is completely ignored, although it plays the crucial role in distinguishing quantum theory from e.g. classical electromagnetism (see
 \cite{NL2}).      By this principle quantum measurements generate discrete events corresponding to interaction of individual quantum systems, say photons or electrons, with measuring devices. Such discrete events are clicks of photo-detectors or points on the screen with photo-emulsion in the diffraction experiments. Bohr call them phenomena. For him, only phenomena can be considered as ``elements of reality''.
We now cite Bohr: 
 
{\it  ``I advocate the  application of the word {\bf phenomenon} exclusively to refer to the observations obtained under specific circumstances including an account of the whole experimental arrangement. In such terminology, the observational problem is free of any special intricacy since,  in actual experiments, all observations are expressed by
unambiguous statements referring, for instance, to the registration of the point at which
an electron arrives at a photographic plate. Moreover, speaking in such a way is just suited to emphasize that the the appropriatephysical interpretation of the symbolic quantum mechanical
formalism amounts only to predictions, of determinate or statistical character,
pertaining to individual phenomena appearing under conditions defined by classical physical concepts.''} (\cite{BR},v. 2, p. 64]

It seems that  Zelinger's principle of information quantization  is just an information reformulation of Bohr's principle of individuality of quantum phenomena.\footnote{  We remark once again that Bohr considered the complementarity principle as an epistemological principle, as a principle about  extraction of information about features of quantum systems with the aid of measurement devices. In fact, Bohr's views match very well with modern development of quantum information theory (see Plotnitsky \cite{PL1,PL2}, Jaeger \cite{Jaeger1,Jaeger2}), including the information interpretation of QM (Zeilinger-Brukner \cite{Z,B1,B2}), QBism \cite{F1,F2}, and information reconstruction of quantum theory \cite{DA1}-\cite{Ch2}.}   

Besides {\bf Individuality}, in the above citation Bohr also emphasized {\bf Contextuality}, {\it an account of the whole experimental arrangement.} We remark that, for to Bohr, {\bf Contextuality principle} is a consequence of {\bf Quantum action principle.} Indivisibility of 
quantum of action implies irreducible dependence of measurement's output on the experimental context. 
Logically {\bf Contextuality}  should generally imply {\bf Context complementarity},  since
the possibility to combine any group of experimental contexts into a single context for join measurement of observables 
is really  surprising. For me, the real surprise is not that some experimental contexts are incompatible, e.g., contexts 
for measurement of position and momentum in QM, but that in some theories, e.g., classical physics, it is assumed mutual compatibility of any pair of contexts.

We complete this section with the remark that in discussions related to violation of the Bell type inequalities the term  ``contextuality'' is used in the very restricted meaning, as dependence on measurement of a compatible observable \cite{Bell1}. In the present paper, as well as in my previous works, e.g., \cite{KHR_CONT}, ``contextuality'' was used to note dependence on a general experimental context (``whole experimental arrangement''). To speak about contextuality, we need not to consider two observables;  we can speak, e.g., about the context of position measurement or the context of measurement of the concrete spin projection. 

\section{Fundamental principles of quantum mechanics}

The above considerations lead to the fundamental principles of QM: 
\begin{enumerate}
\item {\bf Quantum action principle.}
\item  {\bf Bohr's complementarity principle.}
\end{enumerate}
We consider QM as an epistemic theory \cite{W,Harald1,Harald2,PARX,OE1}, a theory about extraction of knowledge about nature; in terminology of Hertz and Boltzmann this is observational theory \cite{HER,BZ1,BZ2,OE2}. So, these two principles provide the epistemic foundations of quantum theory. 

The quantum action principle is the direct consequence of the quantum postulate (the ontic principle about nature as it is), the second quantum principle (complementarity) is based on the first quantum principle. But their interrelation is complicated 
(see \cite{NL0B,NL1,GG}). 

Of course, these principles  do not provide  representation of QM as a closed formal system. However, in spite of a rather common opinion, even Einstein's relativity based on the principle of relativity and constancy of light's velocity cannot be treated as such a closed system, see Einstein's own comment on this issue (citation is taken from book \cite{Ref}): 

\footnotesize{{\it ``The principle of relativity, or, more exactly, the principle of relativity together with the principle of the constancy of the velocity of light, is not to be conceived as a ``complete system,'' in fact, not as a system at all, but merely as a heuristic principle which, when considered by itself, contains only statements about rigid bodies, clocks, and light signals. It is only by requiring relations between otherwise seemingly unrelated laws that the theory of relativity provides additional statements.''}}

\section{Concluding remarks}

First of all, we emphasize that 
\begin{itemize}
\item EPR-paper \cite{EPR} was not directed against the Heisenberg uncertainty 
and Bohr complementarity principles;
\item Bell's works, see, e.g.,  \cite{Bell0}-\cite{Bell2} and further works in Bell's paradigm, e.g., \cite{CHSH}, were  straightforwardly directed against these principles.
\end{itemize}
However, Bell believed \cite{Bell0} that he is in one boat with EPR. And this belief spread throughout the quantum community.

The recent years were marked by the tremendous success of experimentalists performing the Bell type tests \cite{Hensen}-\cite{Shalm}. 
In the light of this paper (as well as \cite{NL1}), these tests can be considered as the excellent confirmation of the validity of the Bohr
complementarity principle. They also confirmed that the correlations  predicted withing the quantum theory can be preserved at long distances. In this paper, we do not try to provide ``deeper explanation'' of these correlations than given by the quantum formalism
(see a short remark at the very end of Appendix 1). We just wanted to point that the attempt of their ``explanation'' in the Bell framework was suspicious  from the very beginning (i.e., without derivation of any inequality), as an attempt to disprove the complementarity principle.  

The ontological seed of the complementarity principle is the quantum action postulate. Therefore rejection of complementarity is impossible without rejection of the existence of indivisible quantum of action. 

Following \cite{Z}, we searched for the fundamental principles of QM. These are two principles, the quantum action and complementarity principles. The first principle is the epistemological representation of the quantum action postulate.

Finally, I conclude that if the quantum foundations are presented as in section \ref{ph1}, i.e., similarly to the foundations of special relativity,  then {\it the attempts to go beyond the complementarity principle, e.g., with hidden variables of the Bell type, can be compared with the attempts to go beyond special relativity, by rejecting Einstein's  principle on constancy of light's velocity.}

\section*{Appendix 1: Two faced Janus of quantum nonlocality}

The first time L\"uders nonlocalty  was briefly mentioned in EPR-paper \cite{EPR} as the absurd alternative to incompleteness of QM. During the Einstein-Bohr debate non of the debaters considered this alternative seriously. Unfortunately, Einstein 
mentioned nonlocality at a few other occasions and highlighted it in \cite{EB} with the shicky slogan, ``spooky action at a distance.'' This sort of nonlocality is the straightforward 
consequence of using the projection postulate in combination with the  Copenhagen interpretation: the quantum state is the state of the individual quantum system. We remark that the Copenhagen interpretation has many versions. 
Plotnitsky even proposed to speak about the interpretations in the spirit of Copenhagen. We characterize such interpretations by emphasizing the individual character of a state. The alternative interpretation is the statistical or ensemble interpretation. Here the
quantum state characterizes the features of an ensemble of identically prepared quantum systems.

{\bf L\"uders nonlocalty:} The state update as back-action of measurement is mathematically formalized by the L\"uders  projection postulate. For a compound system $S=(S_1,S_2),$ measurement on $S_1$ with the concrete output $A=a$  ``instantaneously'' modifies the state of $S_2.$ Here, the crucial role is played by a meaning of   ``instantaneously''. In what space?
If one follows the individual interpretation of the state, then this instantaneous change happens in physical space. 
One really can imagine that this instantaneous change is a consequence of spooky action at a distance. 

However, as was explained in very detail in \cite{NL3}, if one uses the statistical interpretation of a quantum state, then 
``instantaneous'' is related not to physical space, but to information space. There is nothing special in ``instantaneous'' change of information. The same happens 
in process of probability update in classical probability theory. Here states of random systems are represented by probability measures. One also might say that such state changes instantaneously.  But, nobody describes this situation as nonlocality.

\medskip

{\bf Bell nonlocality.} This is nonlocality of of some subquantum models invented by Bell and known as models with hidden variables 
\cite{Bell0}-\cite{CHSH}.
The existence of such models is not surprising at all, human imagination is powerful and it can generates a variety of mathematical structures that have nothing to do with physics. How does one couple Bell nonlocality with quantum physics? Bell proposed to compare 
correlations described by subquantum models with quantum correlations, theoretical and experimental. 
As was pointed out in \cite{NL3}, the Bell project does not take into account the ontic-epistemic structure of scientific theories.
Already Hertz \cite{HER} and Boltzmann \cite{BZ1,BZ2}  (and later Schr\"odinger \cite{DAG}) emphasized this difference: theoretical (causal) vs. observational models (see also \cite{OE1,OE2}). Bell tried to identify outputs of the two descriptions. (This approach 
was strongly criticized by De Broglie \cite{DB}. It seems that he was not aware about the works of Hertz, Boltzmann, Schr\"odinger.
However, their views coincide.)   

As was shown in \cite{NL1},  the Bell type inequalities can be considered in the purely quantum framework, as inequalities for correlations described by quantum theory. In this framework, violation vs. satisfaction of these inequalities is equivalent to local incompatibility vs. compatibility of quantum observables. Hence,    paper \cite{NL} demonstrated that these inequalities are statistical tests  for the Bohr complementarity principle (in particular, the Heisenberg uncertainty principle). 

\section*{Appendix 2: Long distance correlations}

QM endowed with the statistical interpretation does not suffer from L\"uders nonlocality and by operating in the purely quantum framework  one can completely ignore Bell nonlocality. However, my opponents say that you cannot operate in the purely quantum framework, that quantum physics really cry for subquantum explanation of long distance correlations which cannot be explained in the quantum framework. We repeat that one should be very careful by using 
the word ``explanation''. The only possible scientific meaning of this word is ``construction of a proper mathematical model''. 
But, with this formulation we again meet the meaning problem, now with the word ``proper''. I think that Bohr, Heisenberg, or Fock
considered  QM as a proper model ``explaining'' the long distance correlations with entangled states. However, it seems that Bell  did not  consider the quantum model as satisfactory for ``explanation'' of these correlations. Why? Because of the  acausal character of quantum measurements. He wanted to reestablish causality.    

\section*{Appendix 3: EPR vs. Bell in relation to Heisenberg's uncertainty principle}

As was emphasized already in the abstract of EPR-paper \cite{EPR}, EPR did not question Heisenberg's uncertainty principle: 

\footnotesize{{\it ``In quantum mechanics in the case of two physical quantities described by non-commuting operators, the knowledge of one precludes the knowledge of the other. Then either (1) the description of reality given by the wave function in quantum mechanics is not complete or (2) these two quantities cannot have simultaneous reality. Consideration of the problem of making predictions concerning a system on the basis of measurements made on another system that had previously interacted with it leads to the result that if (1) is false then (2) is also false. One is thus led to conclude that the description of reality as given by a wave function is not complete.''}}

It is clear that EPR cannot even imagine that (2), i.e., Heisenberg's uncertainty principle, is false. For them, it was clear 
that these two quantities (incompatible observables) cannot have simultaneous reality. In Bell's model with hidden variables, 
quantities $A(a, \lambda)$ and $A(a^\prime, \lambda)$ have simultaneous reality. (Here $a$ and $a^\prime$ are orientations 
of Alice's beam splitter.)  One may say that this Bellian reality, a part of local realism, 
differs from EPR-reality, \footnotesize{{\it ``the possibility of predicting it with certainty, without disturbing the system.''}}
One may say that EPR wrote about experimental predictions, but generally  $A(a, \lambda)$ and $A(a^\prime, \lambda)$ are components 
of some mathematical model, just a possible human image of a causal subquantum model. This is the good point. But, Bell did not proceed in this way. He simply identified the values of random variables of hidden variables with outcomes of quantum observables, the real physical observables. The latter couples of Bellian (hidden variables) reality with the EPR (outcome prediction) reality. So,  Bell's introduction of hidden variables contradicts even the statement in the abstract of the EPR-paper ...   

Finally, I remind that this viewpoint was presented a long ago  by De Broglie as his reaction to Bell's inequality  \cite{DB}.  

\section*{Appendix 4: Reestablishing causality}

Can one reestablish causality without contradicting to Heisenberg's uncertainty principle? I think that the answer 
is ``yes'' and the corresponding mathematical model was constructed in a series of my papers, see, e.g., \cite{PCSFT1,Beyond,PCSFT2}. Of course, such reestablishing 
cannot be done in such a trivial way as in the Bell model with hidden variables, i.e., though simple identification of the values 
of functions of hidden variables with experimental outcomes. 

In \cite{PCSFT1,Beyond,PCSFT2}, I developed {\it prequantum classical statistical field theory} (PCSFT), reproducing quantum probabilities 
and correlations within theory of  classical random fields. PCSFT is a kind of hidden variables model, but the values of classical random variables, functions of  classical random fields, are not identified with the outcomes of quantum observables. The PCSFT-counterpart of a quantum observable which is
represented by Hermitian operator $A$ are given by quadratic form $f_A(\phi)= \langle \phi \vert A\vert \phi\rangle.$ The range 
of values of $f_A$ does not coincides with the spectrum of $A.$ In particular, if $A$ has the spectrum $\{-1,+1\},$ 
the range of values of $f_A$ is not bounded by 1. Correlations of such quadratic forms can violate the Bell-type inequalities.

PCSFT is a causal theoretical model for the observational model, QM (see Hertz and Boltzmann \cite{HER,BZ1,BZ2}, 
see also \cite{OE1,OE2}). PCSFT can be straightforwardly  connected with observations 
through mapping onto QM.  However, If one is looking for causal coupling with observations, then PCSFT has to be endowed with 
its own observation theory. Such a theory should describe generation of measurement outputs from quadratic forms 
$\phi \to f_A(\phi).$ The first steps towards such PCSFT-based measurement theory were done in \cite{PCSFT2}. This measurement theory 
is based on detectors of the threshold type. It does not violate the Heisenberg uncertainty nor the Bohr complementarity 
principles. (However, the role of indivisible quantum of action in this theory has not yet been clarified.) In particular, the PCSFT-generated observational model reproduces violation of the CHSH inequality for discrete clicks of
 detectors (of the threshold type) \cite{Beyond}. This model has nontrivial coupling with the temporal structure of measurements, in particular, their constraining by time-coincidence of detections for Alice's and Bob's detectors.

 \end{document}